\documentclass[letter]{aa}
\usepackage[utf8]{inputenc}
\usepackage[T1]{fontenc}
\usepackage{amsmath}
\usepackage{amssymb}
\usepackage{capt-of}
\usepackage{hyperref}
\usepackage{mlmodern}
\usepackage[babel]{microtype}
\usepackage[svgnames]{xcolor}
\usepackage[varg]{txfonts}
\usepackage{graphicx}
\usepackage{natbib}
\usepackage{flushend}
\usepackage{booktabs}
\usepackage{longtable}
\usepackage{rotating}
\usepackage[normalem]{ulem}
\usepackage{amsmath}
\usepackage{amssymb}
\usepackage{caption}
\usepackage{subcaption}
\usepackage{siunitx}
\usepackage{hyperref}
\usepackage{orcidlink}
\usepackage{natbib}
\bibpunct{(}{)}{;}{a}{}{,} 
\hypersetup{
    colorlinks=true,
    citecolor=blue,
    urlcolor=blue,
    linkcolor=blue,
}

\titlerunning{LyC escape from tidal bridge in LACES 104037}
\authorrunning{T. E. Rivera-Thorsen et al.}
\date{}
\title{Lyman Continuum escaping from in-situ formed stars in a tidal bridge at \(z=3\)}
\hypersetup{
 pdfauthor={T. Emil Rivera-Thorsen},
 pdftitle={Lyman Continuum escaping from in-situ formed stars in a tidal bridge at \(z=3\)},
 pdfkeywords={},
 pdfsubject={},
 pdfcreator={Emacs 30.2 (Org mode 9.8-pre)},
 pdflang={English}}
\begin{document}

\author{
 T. E. Rivera-Thorsen\orcidlink{0000-0002-9204-3256}\inst{1}\fnmsep\thanks{Corresponding author. \email{trive@astro.su.se}}
\and
A. Le Reste\orcidlink{0000-0003-1767-6421}\inst{2}
\and
M. J. Hayes\orcidlink{}\inst{1}
\and
S. Flury\orcidlink{0000-0002-0159-2613}\inst{3}
\and
A. Saldana-Lopez\orcidlink{0000-0001-8419-3062}\inst{1}
\and
B. Welch\orcidlink{0000-0003-1815-0114}\inst{4}
\and
S. C. Choe\orcidlink{0000-0003-1343-197X}\inst{1}
\and
 K. Sharon\orcidlink{0000-0002-7559-0864}\inst{5}
\and
K. Kim\orcidlink{0000-0001-6505-0293}\inst{6}
\and
M. R. Owens\orcidlink{0000-0002-2862-307X}\inst{7}
\and
E. Solhaug\orcidlink{0000-0003-1370-5010}\inst{8, 9}
\and
H. Dahle\orcidlink{0000-0003-2200-5606}\inst{10}
\and 
J. R. Rigby\orcidlink{0000-0002-7627-6551}\inst{11}
\and 
J. Melinder\orcidlink{0000-0003-0470-8754}\inst{1}
}
\institute{
 The Oskar Klein Centre, Department of Astronomy, Stockholm University, AlbaNova, 10691 Stockholm, Sweden
\and
Minnesota Institute for Astrophysics, University of Minnesota, 116 Church Street SE, Minneapolis, MN 55455, USA
\and
Institute for Astronomy, University of Edinburgh, Royal Observatory, Edinburgh, EH9 3HJ, UK
\and
International Space Science Institute, Hallerstrasse 6, 3012 Bern, Switzerland
\and
Department of Astronomy, University of Michigan, 1085 S. University Ave, Ann Arbor, MI 48109, USA
\and 
IPAC, California Institute of Technology, 1200 E. California Blvd., Pasadena CA, 91125, USA
\and
Department of Astronomy, University of California, Berkeley, Berkeley, CA 94720, USA
\and
Department of Astronomy and Astrophysics, University of Chicago, 5640 South Ellis Avenue, Chicago, IL 60637, USA
\and
Kavli Institute for Cosmological Physics, University of Chicago, Chicago, IL 60637, USA
\and
Institute of Theoretical Astrophysics, University of Oslo, P.O. Box 1029, Blindern, NO-0315 Oslo, Norway
\and
 Astrophysics Science Division, Code 660, NASA Goddard Space Flight Center, 8800 Greenbelt Rd., Greenbelt, MD 20771, USA
}

\abstract { We present an analysis of archival JWST NIRSpec IFS and HST imaging
  observations of the \(z = 3\) Lyman-Continuum Emitter (LCE) candidate
  LACES104037. We show that a nearby galaxy, denoted LACES104037-S, has a
  redshift offset from the main galaxy by only \(\sim 450 \text{ km s}^{-1}\).
  Together with the identification of tidal bridge features between the
  galaxies, this indicates that the galaxies are interacting and most likely in
  the early stages of a merger. We show that the rest-frame LCE cluster sits in
  a tidal bridge towards the companion, about 2.7 kpc from the galaxy's core.
  It is faint in non-ionizing stellar continuum, and shows faint but
  non-negligible H\(\alpha\) and [\ion{O}{iii}] emission, suggesting that much
  of the gas surrounding the LCE cluster has been dispersed by feedback in the
  shallower gravitational potential of the tidal bridge. From comparing the
  direct LyC escape and the local H\(\alpha\) emission, we find a total
  ionizing escape fraction of \(f_{\text{esc}}^{\text{LyC}} = 57 \pm 8\%\) from
  the LCE cluster. We estimate the age of the LCE cluster to be \(\lesssim
  6.5\) Myr, indicating that the cluster must have formed in situ in the tidal
  bridge well after the time of closest interaction. LyC escape from tidal
stripping or in-situ formed stars in tidal features would depend less on
intrinsic galaxy properties than typically observed in low-\(z\) LCE surveys,
and could help explain the higher cosmic escape  fraction as well as the
enhanced diversity of LCE galaxy properties observed at Cosmic Noon.}

\keywords{ Galaxies: kinematics and dynamics---Galaxies: starburst---Galaxies:
star formation---Galaxies: interactions---Galaxies: individual: LACES104037 }

\maketitle \nolinenumbers \section{Introduction} \label{sec:orgfdef85c} Young,
star-forming galaxies  contributed a  major part  of the  ionizing
Lyman-Continuum (LyC) photons that re-ionized  the intergalactic medium (IGM)
in the Epoch  of   Reionization  (EoR),  the  last   major  phase  transition
of   the  Universe \citep[e.g.,][]{fauchergiguere2020, haardt2012, wise2019}.
Estimates of the cosmic LyC escape fraction required to account  for the EoR
are centered around  10--20\%
\citep{robertson2015,naidu2020,giavinazzo2025arxiv}; some    estimates go
higher \citep{davies2021}, while some go as low as \(\sim 5\%\)
\citep{finkelstein2019,atek2024}. At low redshifts, the average escape fraction
is consistently  far below these values.  At \(z \lesssim 0.1\) it is negligible,
with only two known Lyman-Continuum  emitter (LCE) galaxies, each  with
\(f_{\text{esc}}\lesssim 5\%\) \citep{puschnig2017,komarova2024}. In the
low-redshift Universe, \(z \lesssim 0.4\), about 70 LCEs are known, some of
which have very high escape fractions \citep[e.g.,][]{borthakur2014,
izotov2018b, izotov2025,flury2022a,malkan2021b}; but  they  are rare  and
the  cosmic escape  fraction  is   very  low. Stacking  analyses  at \(z \sim
1\) and \(z \sim  3\) have consistently shown upper limits to the average
escape fraction well below what is required to account for the  EoR
\citep{cowie2009,rutkowski2016,grazian2017,alavi2020}.  Strong LyC escape
depends on a high production rate of ionizing photons, and on mechanisms
carving out channels through the neutral ISM to enable the free passage of LyC
photons into the IGM \citep[e.g.,][]{trebitsch2017,kakiichi2021}. In  the
low-redshift  Universe, LyC  escape  has  been found  to correlate with
observables  such as the  UV \(\beta\) slope,  reflecting low dust content and
a young stellar  population with  high LyC  production rate
\citep{ji2020,kim2020,flury2022b,flury2025};  with
[\ion{O}{iii}]/[\ion{O}{ii}]   ratio,  which reflects  the ionization  level of
the ISM  \citep{izotov2018b,flury2022a}; with weak neutral absorption features,
reflecting low \ion{H}{i} column density along the line of sight
\citep{chisholm2018a,gazagnes2020, saldanalopez2022}; and with  broad emission
line components  reflecting strong  bulk outflows which  can  rupture the
\ion{H}{i}  envelope  around  the  source star  cluster
\citep{alexandroff2015,chisholm2017,mainali2022,amorin2024}. However, many of
these correlations predicting \(f_{\text{esc}}^{\text{LyC}}\), such as
[\ion{O}{iii}]/[\ion{O}{ii}], \(\beta_{\text{UV}}\), and Ly\(\alpha\)
equivalent width, are weakened or break down at \(z \gtrsim 2\) \citep{zhu2025,
giavinazzo2025arxiv, riverathorsen2022, saxena2022, bassett2022, kerutt2024,
mascia2025, citro2025}, leaving a more diverse population of leakers observed
at higher redshifts. This shift to a more diverse leaker population suggests
that LyC escape at \(z \gtrsim 2\) depends less on intrinsic properties of the
leaking galaxies than at lower redshifts. 

\begin{figure*}[htbp] \centering \includegraphics[width=0.93
  \textwidth]{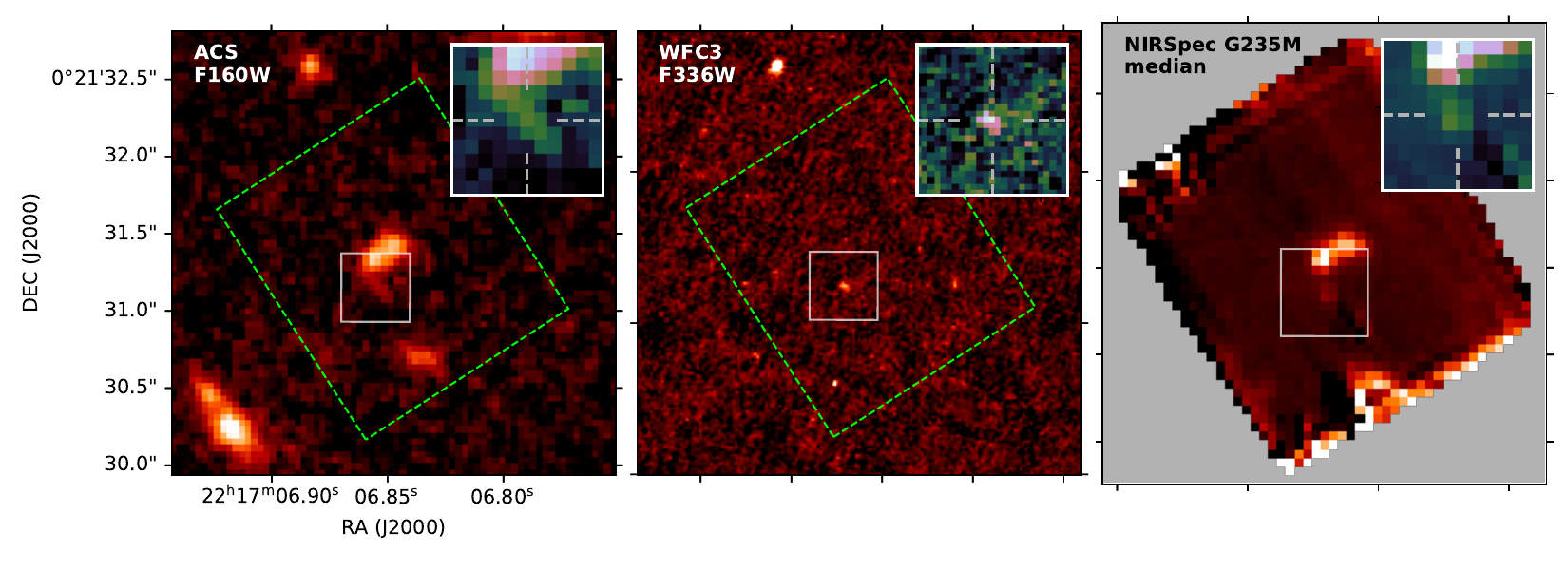}
  \caption{\label{fig:align}Archival Hubble observations in F160W probing the
    rest-frame B band (left) and F336W showing rest-frame LyC (center), along
    with a rest frame optical continuum image of the JWST/NIRSpec IFU cube
    (right), showing LACES104037 and its immediate surroundings. Green dashed
    squares in the HST images show the approximate footprint of the
    JWST/NIRSpec observations. Insets show a zoom in on the LACES104037 itself,
    centered on the LCE.} 
\end{figure*}

Mergers and major interactions have been suggested  by multiple authors as
triggers of LyC escape, as they can both induce strong star formation and
displace stars from gas \citep[e.g.,][]{bridge2010, bergvall2013, lereste2024};
and  indeed, the  cosmic rate  of  mergers and  major interactions has been
found to rise by an order of magnitude  from \(z \sim 0.3\)  to \(z \sim 3\),
after which it stays within a factor  of a few, roughly consistent with the
evolution in \(f_{\text{esc}}^{\text{LyC}}\) \citep{puskas2025,duan2025}.
Indeed, in the \(z \sim 0.3\) Low-Redshift  Lyman  Continuum  Sample, follow-up
imaging  has shown  that at least 40\%  of these galaxies  show signs  of
undergoing mergers \citep{lereste2025barxiv}, a strong  over representation
compared to the \(\sim5\%\) of the general low-redshift population
\citep[e.g.][]{darg2010}. \cite{kostyuk2025} found that simulated galaxies at
\(6 \lesssim z \lesssim  10\) had  an average  \(f_{\text{esc}} \sim 15\%\)  if
they  were merging, compared  to \(\sim  3\%\)  if they were not.
Observationally, \cite{zhu2025}  found from  visual classification  that 20  of
23  LCEs at  \(z \sim  3\) in GOODS-South  were undergoing  mergers, and
\cite{yuan2024} found that a dominant fraction of \(z \gtrsim 3\) LCEs show
spatial offsets between LyC emission and non-ionizing continuum. In contrast,
\cite{mascia2025} compared  the merger rate  of a number  of EoR galaxies  to
statistical tracers of LyC escape  and found no  statistical correlation;
however, these authors relied on LCE tracers calibrated  in the local Universe.
While they  did adjust for known cosmic  evolution of  these tracers, such an
analysis is  still vulnerable to  any unknown biases  built-in to these  local
calibrations, biases  which are evident  from the increased diversity of LCEs
observed at \(z \gtrsim  3\). Mergers can  cause LyC escape through either
induced bursty star formation or  spatial displacement  of neutral gas from the
bulk of the galaxy's stars. Merger-induced central starbursts  are not
qualitatively different from other starbursts, and thus should  not give rise
to any deviation  from the correlations between LCE and bursty  star formation
established at low redshifts. In contrast, tidal displacement of neutral ISM
from the  bulk of the stars and possible in-situ star formation in outer
regions of a galaxy, leads to less dependence on intrinsic galaxy properties of
LyC escape, and can potentially give rise  to the observed diversity of LCE
galaxies observed at \(z \gtrsim  2\). However, such scenarios have barely been
studied observationally. One example of LyC escape facilitated by tidal
stripping of \ion{H}{i} during a major merger was observed directly in the
nearby galaxy Haro 11 by \citet{lereste2024,ejdetjarn2025}.
\cite{riverathorsen2025arxiv} reported circumstantial evidence for a similar
scenario in the gravitationally lensed \(z = 2.37\) Sunburst Arc. In this
work, we present observational  evidence for Lyman Continuum  escaping from
young stars formed  in situ in a  tidal bridge extending from  a starburst
galaxy at  \(z = 3\), towards a  spectroscopically confirmed  interacting
companion. This adds to the mounting evidence that such configurations might be
more common than previously thought. Throughout this paper, we assume a
standard flat \(\Lambda\)CDM cosmology with \(H = 70 \text{ km
s}^{-1}\text{Mpc}^{-1}\) and \(\Omega_{\text{M}} = 0.3\). Images are oriented
North up, East left.

\section{Target and observations}
\label{sec:orgb21b58b}

LACES104037  was  first  reported  as  an LCE  by  \citet{fletcher2019}.  These
authors classified it  as part of  their ``silver'' sample because  the offset
between  the observed rest-frame LyC and the core of non-ionizing continuum was
slightly  more than 0.6'', which they associate with an elevated risk of
interloper contamination. It was selected as a  target for NIRSpec IFU
observations  in Cycle 1 (PID 01827; PI: Kakiichi) for its stellar morphology,
which is among  the more extended and complex in the sample of
\cite{fletcher2019}. The JWST observations were carried out on October 27th
2022, using NIRSpec in IFU  mode with the F170LP/G235M filter/grating
combination. We  downloaded the Level  3 spectral  cubes from MAST. We have
not attempted any further  reduction or cleaning steps than  what have been
done  by  the automated  pipeline,  except  for  manually  aligning and
subtracting  the background cube. We also obtained archival HST imaging data
from the LymAn Continuum Escape Survey (LACES, PID:  14747, PI:  Robertson),
taken with WFC3/F336W and  ACS/F160W. These data are described by
\citet{fletcher2019}. For UVIS/F336W, we obtained the Hubble Advanced Products
- Multi Visit Mosaic (HAP-MVM) products from MAST, which offered a convenient
high-quality data product. We obtained the combined ACS/F160W frames from the
Hubble Legacy Archive (HLA). The WCS systems  were well aligned in F336W  and
NIRSpec already, but was  somewhat off in F160W. We used a number of compact
sources in the field to make sure F160W was aligned to the other observations
to within \(\sim 0\farcs05\).

\section{Analysis and results}
\label{sec:org3994819}
\subsection{Stellar morphology}
\label{sec:org1023e5a}

LACES 104037 and surroundings as seen in these observations are shown together
in figure \ref{fig:align}.  The left panel shows rest-frame $B$ band F160W
imaging, the center panel shows rest-frame LyC F336W imaging, and the right
panel shows the wavelength-median  image of the NIRSpec cube. The  NIRSpec
footprint is shown  in green in  the HST image panels; the inset axes show
zoom-ins centered on the site of LyC escape. F160W shows  a galaxy with  two
bright cores separated  by \(0 \farcs 3\) along  the SE-NW direction, and  an
apparent tidal tail protruding  from its SE edge, pointing toward another
galaxy to the S at the edge of  the NIRSpec footprint. The center  panel shows
that  the LyC emission is  spatially coincident with  the apparent tidal tail,
rather than the main star forming clumps in the central regions of the galaxy.
The LyC centroid is offset from the mid point between the two central knots by
\(\approx 0\farcs62\).

\subsection{Emission line mapping} \label{sec:org2cbd37d} \begin{figure*}[htbp]
  \centering
  \includegraphics[width=0.995\textwidth]{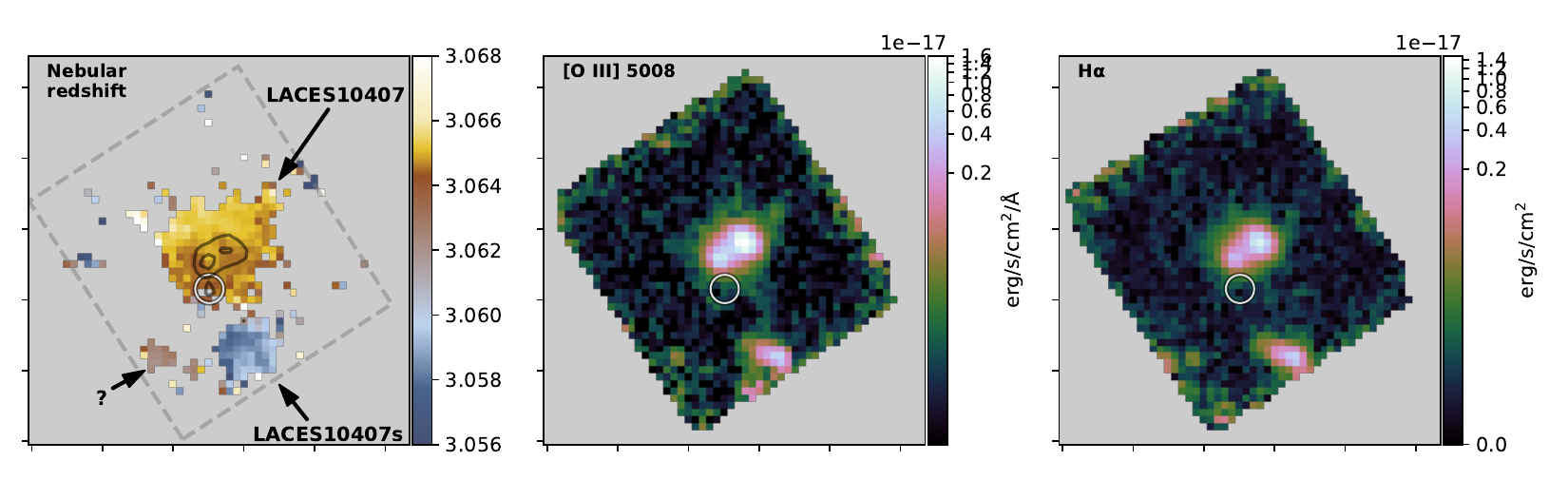}
  \caption{\label{fig-linesfig}Kinematics and [\ion{O}{iii}] and H$\alpha$ line
    emission in LACES104037 and the companion LACES104037-S. The Redshift is
    masked to only include spaxels where S/N([\ion{O}{iii}]) > 3 and the line
    fitting successfully converged on a solution. This map shows that
    LACES104037-S is close to LACES104037 in velocity space and most likely an
    interacting companion. The circle shows the approximate location of the
    escaping LyC, and the black contours in the redshift map show stellar
    continuum levels from the right panel in \autoref{fig:align}. The question
    mark in the left panel points to a faint possible third interacting companion.}
\end{figure*}

We performed line fits of the emission lines of LACES104037 in each spaxel of
the NIRSpec IFU cube following  the procedure described in
\citet{riverathorsen2025arxiv}, which we refer to for a more detailed
description.  In short, we initially subtracted the continuum in each  spaxel
by masking  out the strong emission  lines, subsequently taking  a running
median in  a 31 pixel  window, interpolating this onto  the masked pixel,  and
subtracting this   running  median.   We   then   used  the   line   fitting
software   \texttt{CubeFitter.jl} \citep{riverathorsen2025cubefitter} to
simultaneously fit the lines H\(\beta\), [\ion{O}{iii}]\(\lambda\lambda\)
4960,5008, H\(\alpha\), and [\ion{N}{ii}]\(\lambda \lambda\) 6548,6584  to
single Gaussian profiles with shared redshift  and line width. At a resolving
power of \(R \sim 1000\), G235M does not resolve the  line widths of the
relevant lines. In addition to the line fits, we have also computed line fluxes
by numerical integration over a window 2000 km s$^{-1}$ wide centered around
the redshift $z_0 = 3.062$, selected to cover the full line in both component
galaxies. \autoref{fig-linesfig} shows the resulting kinematic and  line
emission  maps. Importantly, the fitting shows that the visual companion galaxy
to the  S of  LACES104037 is indeed  a physical,  interacting companion,
separated from  LACES104037 by a  projected distance  of \(\sim12.5\) kpc
(approximately 1.5 \(\times\) the  radius of  the Milky  Way disk);  and
blueshifted  by a  relative line-of-sight velocity of  \(\sim450\) km
s\(^{-1}\). We found no designation for this companion in NED, so here we
designate it LACES104037-S. We also note a small group of spaxels to the SE
with redshift in between the two major galaxies; possibly a third, minor
interacting party. The right panel shows  the  [\ion{O}{iii}]\(\lambda\) 5008
numerical line flux map.  This map shows a faint, continuous bridge of ionized
gas between LACES104037 and the companion galaxy, and significant emission from
both. We note that the LyC escape is not coincident with locally  elevated
emission strength  in  [\ion{O}{iii}]. Stars and gas are often separated in
tidal features \citep{hibbard2000}; so plausibly, the amount of gas in the
tidal bridge is small and relatively easy for stellar wind from young stars to
have cleared away. If this is a common occurrence in the \(z \gtrsim 2\)
Universe, it could constitute a significant source of ionizing radiation not
well accounted for by low-redshift calibrations.

\subsection{Ionizing escape fraction and stellar population age} We estimated a
local, omnidirectional \(f_{\text{esc}}^{\text{LyC}}\) from the LCE clump by
comparing the H\(\alpha\) and LyC flux. We extracted a spectrum from the
\(2\times2\) IFU spaxels seen in the cross-hairs in the right panel of
\autoref{fig:align}, approximately centered on the LCE, measured the
H\(\alpha\) line flux, and used the conversion factor from \cite{kennicutt1998}
to translate this into a ionizing photon flux \(q(H^0)\). From visual
inspection of the NIRSpec IFU PSF\footnote{Using \texttt{stpsf},
{https://stpsf.readthedocs.io}}, we found that roughly 40\% of the H\(\alpha\)
emission in the LCE is contributed by the PSF of the main galaxy, while the
aperture loss of a point source in a 2x2 aperture is 40\%, making the two
effects approximately cancel each other. We added an estimated 15\% uncertainty
to the measured \(F(H\alpha)\) to account for these unknowns. We measured the
LyC flux in a matching aperture, and estimated the photometric errors as the
standard deviation of 500 randomly placed apertures on empty nearby regions. We
converted LyC flux to an ionizing photon flux using the photon energy at 825 Å,
assuming that the redshift-corrected central wavelength in F336W is also the
average wavelength of LyC photons from this source. We found a
\(F(\text{H}\alpha) = (4.7\pm0.2) \times 10^{-19}\)~erg~cm\(^{-2}\)~s\(^{-1}\)
yielding \(q(\text{H}\alpha) = (2.7 \pm 0.1) \times
10^{-7}\)~cm\(^{-2}\)~s\(^{-1}\); and \(F(\text{LyC}) = (8.7 \pm 1.9) \times
10^{-18}\)~erg~cm\(^{-2}\)~s\(^{-1}\) yielding \(q(\text{LyC}) = (3.62 \pm
0.79)\times 10^{-7}\)~cm\(^{-2}\)~s\(^{-1}\). Assuming no significant dust
extinction and no LyC flux outside the wavelength range of the filter, this
leads to \(f_{\text{esc}}^{\text{LyC}} = 0.57 \pm 0.08\). We also estimated
\(W(H\alpha)\) in this spectrum, taking into account the ionizing escape
fraction. We found an approximate redshift- and LyC escape-corrected rest-frame
\(W(H\alpha) \approx 340\pm50\)~Å. Comparing this to instantaneous burst S99
models\footnote{{https://massivestars.stsci.edu/starburst99/figs/fig83.html}}
with metallicities between 20\% and 200\% \(Z_{\odot}\), we consistently find
an approximate stellar ages of \(6 \pm 1\)~Myr, across IMF slope and mass
cutoff.

\subsection{Galaxy interaction timescale}

The  two interacting  galaxies  have redshifts  of  \(z_{\text{LACES104037-S}}
\approx  3.059  \text{ and  } z_{\text{LACES104037}} \approx 3.065\), yielding
a relative  LOS velocity component between the two galaxies of  \(v_{\parallel}
=  v_{\text{rel}} \times  \cos \theta  \approx 445 \text{  km s}^{-1}\),  where
\(\theta\)  is the viewing angle. The transverse physical distance component
between the galaxies is \(d_{\perp} = d  \times \sin \theta  \approx 12.5\)
kpc.  From this, the time  since interaction is  given as \(t  = d
v_{\text{rel}}^{-1} = d_{\perp}  v_{\parallel}^{-1}(\tan \theta)^{-1} \approx
27  \text{ Myr } (\tan \theta)^{-1}\).  We do not know the viewing angle of the
relative motion, but reasonable angles of \(30 \text{ and } 60\deg\) correspond
to a time since closest approach of \(t \approx 45 \text{ and } t \approx 15
\text{ Myr}\).  Since the stellar population age is \(\lesssim 6.5 \)~Myr, it
is unlikely that the leaking stars existed already at the time of closest
approach; instead, they have likely formed inside the tidal bridge formed by
the interaction.

We have also considered the possibility that the dual cores of LACES104037
might be a sign that the galaxy is itself a late-stage merger, and that the
tidal tail might instead originate from this interaction. However, while the
separation from LCE to the nearest core is only \(0\farcs 36\), the velocity is
\(v \lesssim 80 \text{ km s}^{-1}\), leading to a travel time of \(t \approx 34
\text{ Myr} \times (\tan{\theta})^{-1}\)~Myr; still favoring in-situ formation.

\section{Summary}

In this work, we have found that the LCE candidate galaxy LACES104037 from the
work of \cite{fletcher2019} is interacting with a nearby galaxy, LACES103037-S;
and that the candidate LyC emission reported in that work originates from
in-situ formed massive stars in a tidal feature towards the companion. We find
that LACES104037 differs from typical LCEs at low redshift in that the escaping
LyC originates, not from the bright and highly star forming core, but from an
in-situ formed star cluster in a tidal bridge toward the companion, offset from
the center of the galaxy by a projected distance \(\approx 4.8\)~kpc, in a
region not associated with elevated nebular line emission. We find traces of a
possible third, smaller companion system to the SE of the main companion
galaxy, with a redshift intermediate to the two larger galaxies. We find faint
but definite, extended line emission tendrils and halos surrounding both the
major galaxies, marginally overlapping between them. Such significant LyC
escape from the very outskirts of a galaxy suggests that a---potentially
significant---sub-population of LCE galaxies, particularly at \(z \gtrsim 2\),
might not be well characterized by the \(z \lesssim 0.5\) samples typically
used to calibrate observational \(f_{\text{esc}}^{\text{LyC}}\) proxies. Many
LCE surveys select against an LCE like LACES104037 by excluding morphologically
complex galaxies and candidates with LyC emission offset from the main
non-ionizing continuum emission, based on an elevated risk that such systems
are contaminated by foreground interlopers. Such caution is well founded, as it
can be difficult to distinguish morphologically complex systems from
interlopers; but works such as this one show that it might also lead to
exclusion of a significant number of LCEs and consequently an underestimation
of the diversity of Lyman-Continuum emitters. 

\begin{acknowledgements} This work is based on observations made with the
  NASA/ESA/CSA James Webb Space Telescope and the NASA/ESA Hubble Space
  Telescope. The data were obtained from the Mikulski Archive for Space
  Telescopes at the Space Telescope Science Institute, which is operated by the
  Association of Universities for Research in Astronomy, Inc., under NASA
  contract NAS 5-03127 for JWST and NAS 5-26555 for HST. These observations are
  associated with programs 14747 and 01827. ER-T is supported by the Swedish
  Research Council grant 2022-04805. 
\end{acknowledgements}

\bibliographystyle{aa}
\bibliography{AllPapers}
\end{document}